\documentclass[conference]{IEEEtran}
\IEEEoverridecommandlockouts
\usepackage{multirow}
\usepackage{float}
\usepackage{cite}
\usepackage{amsmath,amssymb,amsfonts}
\usepackage{algorithmic}
\usepackage{graphicx}
\usepackage{textcomp}
\usepackage{xcolor}
\usepackage{comment}
\renewcommand\IEEEkeywordsname{Keywords}
\def\BibTeX{{\rm B\kern-.05em{\sc i\kern-.025em b}\kern-.08em
    T\kern-.1667em\lower.7ex\hbox{E}\kern-.125emX}}
\begin{document}

\title{Dualband and Tripleband Metamaterial absorber in WR1 band and Lower TeraHertz frequencies\\

}

\author{\IEEEauthorblockN{Sarvesh Gharat}
\IEEEauthorblockA{\textit{Department of Electronics and Telecommunication} \\
\textit{Vishwakarma Institute of Information Technology}\\
Pune, India \\
sarveshgharat19@gmail.com}

\IEEEauthorblockN{Prutha Kulkarni}
\IEEEauthorblockA{\textit{Department of Electronics and Telecommunication} \\
\textit{Vishwakarma Institute of Information Technology}\\
Pune, India \\
prutha.kulkarni@viit.ac.in}
\and
\IEEEauthorblockN{Shriganesh Prabhu}
\IEEEauthorblockA{\textit{Department of Condensed Matter Physics} \\
\textit{Tata Institute of Fundamental Research}\\
Mumbai, India \\
prabhu@tifr.res.in}

\IEEEauthorblockN{Chandrashekhar Garde}
\IEEEauthorblockA{\textit{Department of Electronics and Applied Science} \\
\textit{Vishwakarma Institute of Information Technology}\\
Pune, India \\
csgarde@gmail.com}
}

\maketitle

\begin{abstract}
Frequency Band of 0.5 THz to 1.1 THz (WR1.5 and WR1) is one of the promising bands when it comes to 6G. In this paper, we propose a novel metamaterial absorber suitable to be used at lower WR1 frequencies in TE mode. Alternatively, the design can be used as a strong absorber at lower Terahertz frequencies. In addition, when used in TM mode the absorber works as a perfect absorber at higher WR1 band.
\end{abstract}

\begin{IEEEkeywords}
Absorber, Kapton, Metamaterials, Terahertz, WR1
\end{IEEEkeywords}

\section{Introduction}

A metamaterial is an artificial material which can achieve electromagnetic properties that are not achievable naturally such as negative refractive index, permittivity and permeability \cite{kshetrimayum2004brief}\cite{capolino2017theory}\cite{walser2000metamaterials}. This happens at the frequencies where the unit periodic cells show resonance \cite{bukhari2019metasurfaces}. A metamaterial absorber is another  interesting application of the same. The first experimentally confirmed metamaterial absorber with 88\% absorption was reported by Landy et al. \cite{landy2008perfect} in 2008. A metamaterial absorber has quite wide applications such as thermal emitters, photovoltaic cells, and optical imaging devices etc to name a few \cite{diem2009wide}
\cite{munday2011large} \cite{hu2010plasmonic}. Various Metamaterial absorbers with wide variety of different properties have been developed and experimentally demonstrated. Many such applications involve modulating the properties by using innovative structural design strategy \cite{zheng2013four}, optimizing the structure, stacking different dielectric layers \cite{zhang2013dual}, or stacking a large number of dielectric layers. \\ 

The metamaterials principally work as an LC circuit whose resonance frequency is inversely proportional to square root of inductance and capacitance \cite{huang2017theory}. Both inductance as well as capacitance depend on multiple parameters such as design and dimensions, period and thickness of substrate used to design the metamaterial. It also depends upon the angle at which electromagnetic field is transmitted. \cite{sangala2020single}\cite{chowdhury2011broadband}\cite{tao2008highly}. There are many attempts made to explain this relation mathematically \cite{marwaha2016accurate}\cite{george2018mathematical}\cite{elander2011mathematical}\cite{bose2012mathematical} but there is no universal formulation for this relation. \\

Since last decade, field of metamaterials has generated tremendous amount of interest among researchers. This is due to it's unique properties and it's applications in multiple fields including public safety, sensor identification, high-frequency battlefield communications, etc \cite{valipour2021metamaterials}\cite{rappaport2019wireless}\cite{singh2015review}. One such application is using metamaterials as an absorber\cite{zhu2018electromagnetic}\cite{tanaka2017metamaterial}. Here, the idea is to minimize the reflectivity by creating a impedance matching condition. This has been experimentally demonstrated at various microwave as well as at terahertz frequency. \cite{landy2008perfect}\cite{tao2008metamaterial}\cite{landy2009design} \\

Metamaterials as an absorber has tremendous amount of application particularly in Gigahertz and Terahertz frequency. Hence, lot of work has been previously done in this field particularly because of need for strong absorbers for thermal detectors at Terahertz frequency \cite{wang2020multiple}\cite{gandhi2021ultra}\cite{liu2020ultrathin}\cite{luo2021multiband}\cite{elakkiya2020seven}.\\

Wang et.al. \cite{wang2020multiple} propose single, dual and triple band absorbers. This has been done by dividing rectangular boundary into multiple parts. The achieved absorbance frequency for all this designs is 2.5 THz for single band, 1.5 and 2.25 THz for dual band and 1.82, 2.83 and 2.93 THz respectively. Gandhi et. al. \cite{gandhi2021ultra} design achieves bandwidth of 3 THz with an absorptivity of more than 90\%. This wide band absorber works in the frequency range of 2.54 THz to 5.54 THz.  \cite{liu2020ultrathin} Liu et. al. have proposed a triband ultrathin design. The authors have analysed their results in both TE and TM mode. The author's have also analysed the change in results after varying multiple parameters like angle, thickness, dielectric of substrate, etc. In \cite{luo2021multiband} Luo et.al. have proposed a CSRR design which has absorbance at both GHz and Terahertz frequency after varying the dimensions. The authors have got central frequencies at 162 GHz, 186 GHz, 1.1 THz and 1.81 THz respectively. In \cite{elakkiya2020seven} Elakkiya et.al. have designed absorber suitable to work in frequency range of 0.3 THz to 0.5 THz. The authors design shows absorbance at 7 different frequencies. This happens to be one of the promising designs in this frequency band.\\

Though all this studies show promising results in their respective frequency bands, it is imperative to have similar absorbers in lower WR1 band.\\

In this study, we propose a novel design for a narrowband metamaterial absorber which can be used in lower WR1 band. WR1 band ranges from 0.75 THz to 1.1 THz \cite{sheikh2021scattering}. The proposed design has resonance frequency at 0.75 THz with absorptivity of 99.42 \%, 1.2 THz with absorptivity of 98.88 \% and 1.28 THz with absorptivity of 96.54\% in TE mode.\\ 

Along with that the design when in TM mode shows absorptvity of 85.37 \% at 0.78 THz and 99.7\% at 1.3 THz. The proposed design has maximum dimensions of $\approx$ 298 $\mu$m and the period after which the unit cell is repeated in 300 $\mu$m. The substrate used is Kapton which has refractive index $\mu$ = 1.88 + 0.04j which is 25 $\mu$m thick.\\ 

To calculate the absorptivity percentage a simple relation between reflection coefficient, transmission coefficient and absorptivity is used $$\text{i.e Absorptivity in }\% = (1 - \text{abs}(S_{11})^2 - \text{abs}(S_{21})^2)\times 100$$ here $S_{11}$ and $S_{21}$ are reflection and transmission coefficients of the form $a + b i$.
Further, because of fully covered perfectly conducting layer at the bottom of the structure, the incident wave cannot be transmitted. Hence, the transmission coefficient can be regarded as zero. Therefore, the deduced equation for absorptivity is
$$\text{i.e Absorptivity in }\% = (1 - \text{abs}(S_{11})^2)\times 100$$
Further, all the simulations done in this study are performed using COMSOL\textregistered Multiphysics v5.6 \cite{multiphysics1998introduction}.

\section{Design, Simulations and Results}

This section includes the proposed design, assumed conditions during simulations and results.

\begin{figure}[hbt!]
  \includegraphics[width=0.5\textwidth]{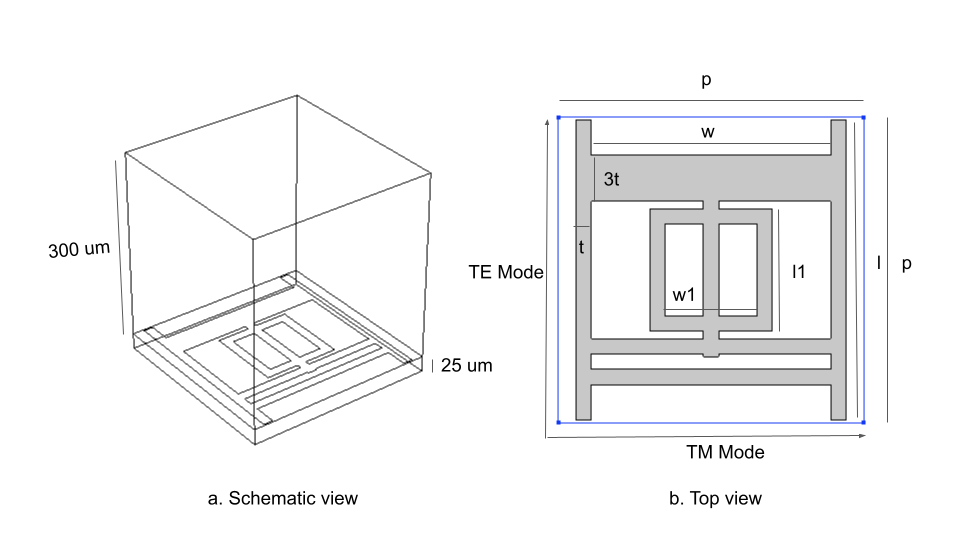}
  \caption{Schematic and Top view of Design}
  \label{fig:fig1}
\end{figure}

Figure \ref{fig:fig1}a shows the schematic of absorber, which has air domain of 300 $\mu$m which is approximately a wavelength at frequency of 1 THz, a rectangular design having symmetry about Y axis but asymmetry about X axis, substrate as a kapton (refractive index: $1.88 + 0.04i$) with thickness of 25$\mu$m and a perfectly electric conducting surface at the bottom. The optimised unit cell is periodically repeated in both X and Y axes with a  period of 300 $\mu$m. \\

Figure \ref{fig:fig1}b shows the front view of unit periodic cell with period(p) of 300 $\mu$m. The proposed design has length(l) of 298 $\mu$m, width(w) of 250 $\mu$m and thickness(t) of 15 $\mu$m respectively. The length(l1) and width(w1) of inner structure is 121 $\mu$m respectively.\\

The proposed design is simulated using COMSOL\textregistered There are certain boundary conditions used while simulating the design.
\begin{itemize}
    \item Perfectly Electric Conductor: PEC is a special case of electric field boundary condition that sets tangential component of electric field to zero i.e n $\times$ E = 0. It is generally used for modelling a lossless metallic surface \cite{multiphysics1998introduction}. In our design this condition is used at the bottom of the substrate and on the structure.
    \item Periodic Condition: Periodic condition: Periodic condition is used to make unit cell structure periodic along particular direction \cite{multiphysics1998introduction}. In our design, unit cell is made periodic along x and y directions.
\end{itemize}

Further, the proposed design is simulated for both TE and TM polarizations. The orientation of the sample with respect to the incident TE and TM polarization of light is shown in Figure~\ref{fig:fig1}~\cite{boardman1991third}.

\begin{table}[ht]
\begin{center}
\begin{tabular}{ |c|c|c|c| } 
\hline
 & Frequency(THz) & Absorptivity(\%)\\
\hline
\multirow{2}{4em}{TM Mode  }& 0.78 & 85.37 \\ 
  & 1.3 & 99.7 \\ 
\hline
\multirow{3}{4em}{TE \hspace{1em} Mode} & 0.75 & 99.42 \\ 
 & 1.2 & 98.88 \\ 
& 1.28 & 96.54 \\ 
\hline
\end{tabular}
\label{table:1}
\end{center}
\caption{Absorptivity \% and it's corresponding frequency in TE and TM Mode}
\end{table}

Table 1 shows the absorptivity and it's corresponding frequency in both TE and TM mode. The difference in absorbance frequency is mainly due to asymmetry kept along Y axis. To understand the mechanism, we have investigated the electric field along z axis at the resonant frequencies.\\

\begin{figure}[hbt!]
  \includegraphics[width=0.5\textwidth]{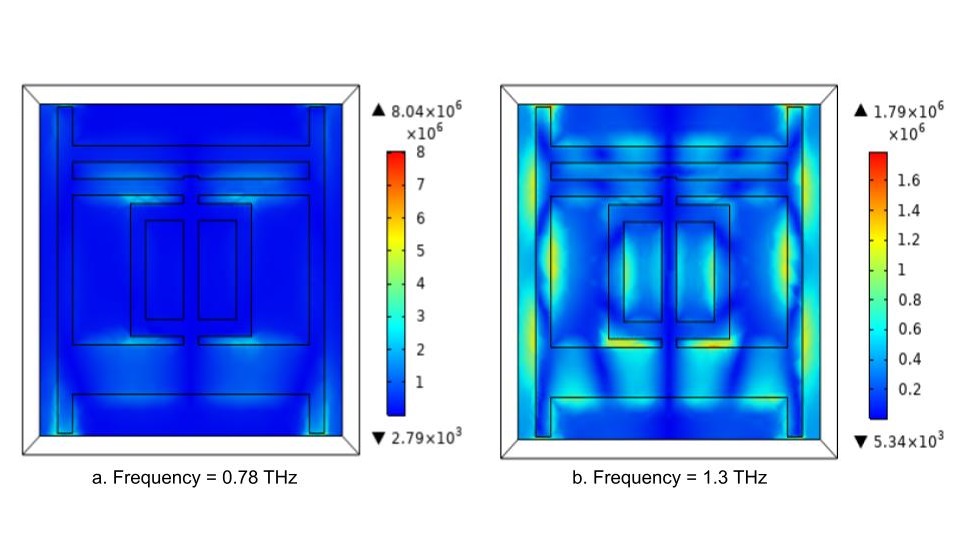}
  \caption{Electric field for resonant frequencies in TM mode}
  \label{fig:eftm}
\end{figure}

\begin{figure}[hbt!]
  \includegraphics[width=0.5\textwidth]{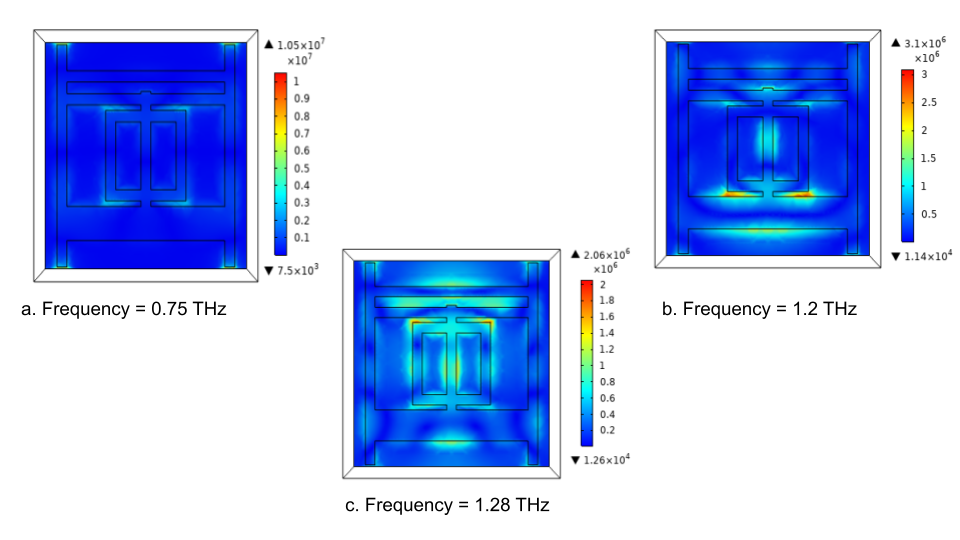}
  \caption{Electric field for resonant frequencies in TE mode}
  \label{fig:efte}
\end{figure}

\begin{figure}[hbt!]
  \includegraphics[width=0.5\textwidth]{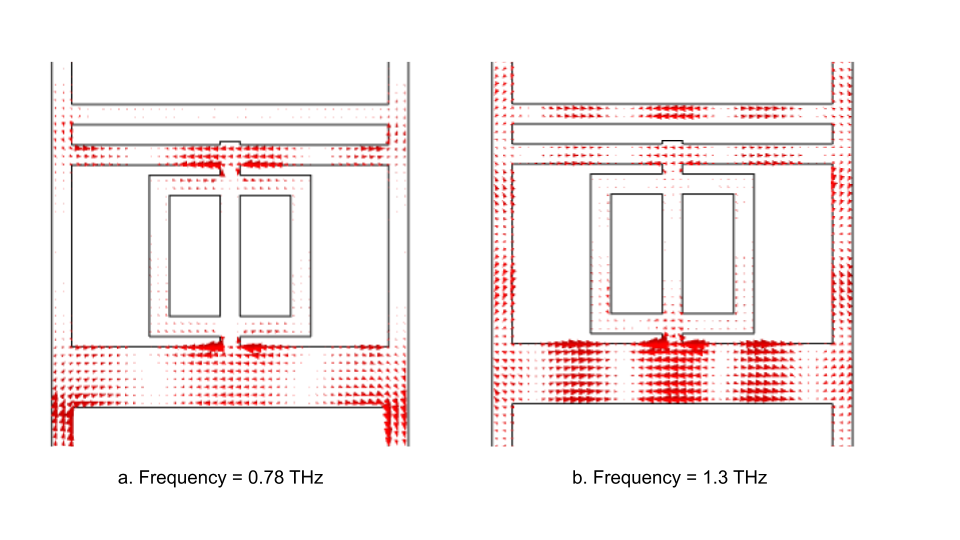}
  \caption{Surface Current Density at resonant frequencies in TM mode}
  \label{fig:scdm}
\end{figure}

\begin{figure}[hbt!]
  \includegraphics[width=0.5\textwidth]{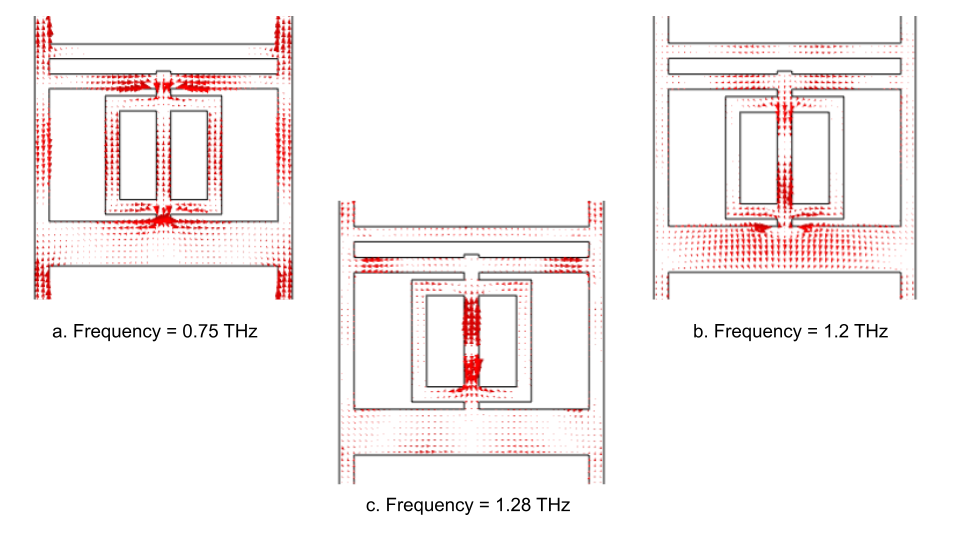}
  \caption{Surface Current Density at resonant frequencies in TE mode}
  \label{fig:scde}
\end{figure}

\begin{figure}[hbt!]
  \includegraphics[width=0.5\textwidth]{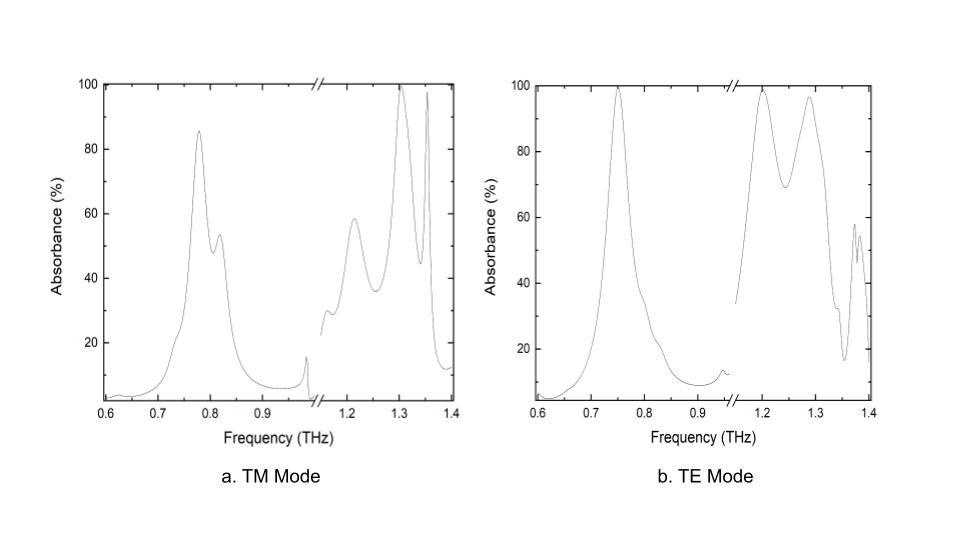}
  \caption{Absorbance vs Frequency graph}
  \label{fig:absorb}
\end{figure}


Figure \ref{fig:eftm}a and \ref{fig:eftm}b shows $E_z$ distribution in TM mode and Figure \ref{fig:efte}a,\ref{fig:efte}b and \ref{fig:efte}c shows $E_z$ distribution in TE mode. It is also important to understand the surface current density distribution on the metamaterial and these are shown in  Figure~\ref{fig:scdm} and Figure~\ref{fig:scde} for TM and TE modes respectively. Along with that Figure \ref{fig:absorb}a,b shows the absorbance vs frequency graph in TM and TE mode from frequency f1 = 0.6 THz to frequency f2 = 1.4 THz. We describe all these the following section.

\section{Discussion}

To find absorptivity at higher WR 1.5 frequency and lower Terahertz frequencies, the proposed design has been simulated in TM mode. Results of which can be seen in Figure \ref{fig:eftm} and Figure \ref{fig:absorb}a. In Figure \ref{fig:eftm}a, we can see the confinement of electric field in the inner part of design which contributes in generating inductance. Apart from that, the electric field can be also see in gap between upper strip and the internal design which contributes in creating capacitance. On the contrary to that at 1.3 THz as evident from Figure \ref{fig:eftm}b, we can see that the electric field is confined in outer part of the structure which again contributes in creating inductance slightly greater than previous case due to increase in strip length. Though, we are unaware about the actual value of inductance and capacitance generated, yet the results show that the confined electric field has generated enough capacitance and inductance to get resonance at these frequencies.\\ 
In Figure~\ref{fig:efte} we show the field distribution of the TE mode. Here, the initial low frequency mode is created by the field getting confined to the upper gaps of the metamaterial wings as seen in Figure~\ref{fig:efte}a. As the THz frequency becomes 1.3~THz, we see that the multimode excitation is taking place and the Field is getting confined to a larger area which results in higher frequency mode. This is seen clearly in Figure~\ref{fig:efte}b. 
Another supporting fact can be found by looking at the current distribution on the Metamaterial. In Figure~\ref{fig:scdm} and Figure~\ref{fig:scde} surface current density distribution is shown for TM and TE modes respectively. From the figure~\ref{fig:scdm} it is clear that the current density shown has several high density regions. These are consistent with the incident polarization of light which is exciting the higher modes in the horizontal arm and there are very weak or negligible excitation on the vertical arms. Especially the center portion of the metamaterial is seen to be having no excitation. The confined high density of currents result in attenuating the emitted radiation. Thus, as the frequency of THz radiation increases, then additional modes are excited in the same arm, preventing the radiation to pass through or get reflected. This leads to those huge absorption peaks seen the spectrum of Figure~\ref{fig:absorb}. \\
For the TE mode excitation, the horizontal arms have almost no excitation modes except the weak ones due to the finite size of the width. However, the center portion of the Metamaterial has a huge strong dipole nature of excitation and this with the dipole excited in the lower arm of the metamaterial, gives rise to two closely spaced absorption resonances as seen in Figure~\ref{fig:scde}b and Figure~\ref{fig:scde}c. These modes are having quite high current density and therefore the incident radiation is well confined, resulting in huge absorption. The TE mode at lower THz frequency can excite across multiple arms of the metamaterial which is seen in Figure~\ref{fig:scde}a where we see one big mode along the frame circling the center portion of the metamaterial. The mode is unidirectional for the three small arms of the center portion which shows that the mode must be large in size which also means that the frequency must be lower.  \\

For achieving near unity absorptivity at lower WR 1.5 frequency and lower Terahertz frequency, the TE mode looks better. Results of which are evident from Figure \ref{fig:efte} and Figure \ref{fig:absorb}b. In Figure \ref{fig:efte}a, we can see that at 0.75 THz, the electric field is mainly confined at the region with gaps which majorly contributes in producing capacitance. Similarly, in Figure \ref{fig:efte}b i.e $E_z$ at 1.2 THz, the electric field is mainly confined in upper part of the gap between internal and external part of design. Apart from that we can also see that the electric field is confined in the upper strip with thickness of 3t producing inductance in the circuit. Similarly at 1.28 THz the confined electric field produces slightly more inductance due to which there isn't a perfect absorbance at this frequency as evident from Table 1 and Figure\ref{fig:efte}b.\\

\section{Conclusion}

We propose a novel design which can act as a perfect absorber at higher WR 1 frequency when used in TM mode and lower WR 1 frequency when used in TM mode. This performance makes the design an ideal candidate to be preferred in this 6G band. Apart from this, irrespective of the mode, the design shows perfect absorptivity at lower Terahertz frequencies i.e At 1.2 THz and 1.3 THz which makes it one of the suitable design to be used as an absorber in thermal detectors.

\section*{Acknowledgement}

The authors would like to that Mr. Ajinkya Punjal (Project Assistant, DCMPS TIFR Mumbai) for his fruitful inputs.

\bibliographystyle{IEEEtran}
\bibliography{example}

\end{document}